\documentclass[final,1p,times]{elsarticle} 
\usepackage{graphicx} 
\usepackage{amssymb} 
\usepackage{amsthm} 
\usepackage{lineno} 

\journal{Nuclear Physics A} 
\begin{document} 

\begin{frontmatter} 


\title{Measurements of soft and intermediate $p_T$ photons from hot and dense matter at RHIC-PHENIX}

\author{Yorito Yamaguchi$^{a}$ for the PHENIX collaboration}

\address[a]{University of Tokyo, 7-3-1 Hongo, Bunkyo, Tokyo 113-0033, Japan}

\begin{abstract} 
The measurements of direct photons in $1.0 < p_{T} < 5.0~$GeV/$c$ for p+p and Au+Au 
collisions have been made through the virtual photon method at PHENIX.
The fraction of the direct $\gamma^{*}$ component to the inclusive $e^{+}e^{-}$ yield 
is determined by a shape analysis using the $e^{+}e^{-}$ mass spectra in 
$m_{ee} < 300~$MeV/$c^{2}$ for $1.0 < p_{T} < 5.0~$GeV/$c$.
The real direct photon spectra in p+p and Au+Au collisions are obtained from the virtual 
direct photon fractions, and a significant excess over a binary-scaled p+p result is seen 
in Au+Au collisions.
Hydrodynamical models which reproduce the Au+Au result indicate the initial temperature of 
the matter is higher than the critical temperature of QGP.
The d+Au data taken in 2008 are promising to evaluate the contribution of the nuclear effects 
due to its large statistics.

\end{abstract} 

\end{frontmatter} 



\section{Introduction}\label{intro}
Direct photons are one of the most powerful probes to investigate properties of the 
matter created by heavy ion collisions since they leave the medium without a strong 
interaction once they are generated.
The direct photons are emitted from every stage of the collisions, and their 
transverse momenta are characterized by their origins.
Especially, the low and intermediate $p_{T}$ region ($p_T \lesssim 5.0$~GeV/$c$) is considered 
as a suitable window for measuring the medium-induced direct photons~\cite{Turbide}.
Primary contributors in the low and intermediate $p_{T}$ region are considered to be 
thermal photons from QGP and photons produced in jet-medium interactions such as 
jet-photon conversions and in-medium Bremsstrahlung, respectively.
It is quite certain that the direct photons below 5.0~GeV/$c$ provide rich 
information of the created matter, but it is very challenging to measure them due to 
a large background from hadron decays.

In the PHENIX experiment, two different analysis methods, namely, a 'real' photon method 
and a 'virtual' photon method, have been successfully developed.
Measurements of 'real' direct photons using an electromagnetic calorimeter in p+p 
and Au+Au collisions have been made successfully in $p_{T} > 4~$GeV/$c$ at the PHENIX 
experiment~\cite{high_pt_pp, high_pt_auau}, and the results are in agreement with 
the next-to-leading-order perturbative QCD (NLO pQCD) calculation.
The measurement using the 'virtual' photon method can be extended to $p_{T} < 4~$GeV/$c$.
The results of the direct photon measurements with the virtual photon method at PHENIX 
are presented in this report.

\section{Virtual Photon Method}\label{vphoton}
In general, any source of real photons can emit virtual photons which convert to low mass 
$e^{+}e^{-}$ pair.
A direct photon production process such as $q+g \rightarrow q+\gamma$ has an associated 
process that produces low mass $e^{+}e^{-}$ pair, i.e. $q+g \rightarrow q+\gamma^{\ast} 
\rightarrow q+e^{+}e^{-}$.
The relation between the photon production and the associated $e^{+}e^{-}$ pair production 
is expressed by Eq.~\ref{eq_KW}~\cite{Kroll-Wada}.
\begin{equation}
\frac{d^{2}n_{ee}}{dm_{ee}} = \frac{2\alpha}{3\pi} \frac{1}{m_{ee}}
\sqrt{1-\frac{4m_{e}^{2}}{m_{ee}^{2}}} \left( 1+\frac{2m_{e}^{2}}{m_{ee}^{2}} \right)
S dn_{\gamma},
\label{eq_KW}
\end{equation}
where $\alpha$ is the fine structure constant, $m_{e}$ and $m_{ee}$ are the masses of the
electron and the e$^{+}$e$^{-}$ pair, respectively, $S$ is a process dependent factor
and $dn_{\gamma}$ is an emission rate of the photons.
In the case of $\pi^{0}$ and $\eta$ Dalitz decays (Kroll-Wada formula), $S$ is given as 
$S = |F(m_{ee}^{2})|^{2} \left( 1- m_{ee}^{2}/m_{hadron}^{2} \right)^{3}$, 
where $F$ denotes the form factor and $m_{hadron}$ is the mass of the parent hadron.
The $S$ factor is obviously zero for $m_{ee} > m_{hadron}$.
On the other hand, the $S$ factor becomes unity for $m_{ee}^{2} \ll p_{T}^{2}$ in the 
case of direct $\gamma^{*}$ decays.
Therefore it is possible to extract the direct $\gamma^{*}$ component from the
e$^{+}$e$^{-}$ mass spectrum by utilizing the difference in e$^{+}$e$^{-}$ mass dependence 
of the $S$ factor.

\section{Measurements in p+p and Au+Au collisions}\label{vphoton_pp_auau}
\begin{figure}[htbp]
  \centering
  \vspace{-0.40cm}
  \resizebox{12.50cm}{!}{\includegraphics{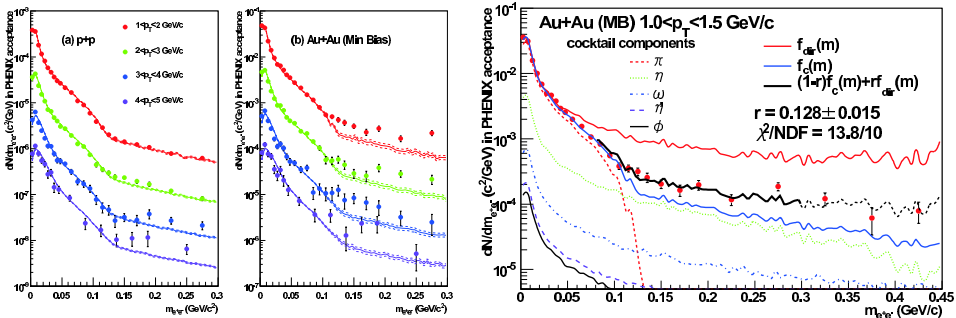}}
  \vspace{-0.5cm}
  \caption[]{The left and center panels show the e$^{+}$e$^{-}$ mass distributions in p+p
    and Au+Au collisions for several $p_{T}$ regions compared to the hadronic cocktail 
    calculations.
    The right panel shows the e$^{+}$e$^{-}$ mass distribution in Au+Au collisions for
    $1.0 < p_{T} < 1.5~$GeV/$c$ together with a fit result by Eq.~\ref{eq_fit}.}
  \label{mass_fit}
\end{figure}

The region $p_{T} > 1.0~$GeV/$c$ and $m_{ee} < 300~$MeV/$c^{2}$ satisfies the requirement 
for applying the virtual photon method that $m_{ee}^{2} \ll p_{T}^{2}$.
The left and center panels in Fig.~\ref{mass_fit} show the e$^{+}$e$^{-}$ mass distributions
in p+p and Au+Au collisions for $m_{ee} < 300~$MeV/$c^{2}$ and $1.0 < p_{T} < 5.0~$GeV/$c$ 
compared to the hadronic cocktail calculations which incorporated the measured yields of the 
mesons at PHENIX.
The symbols and lines show the data and hadronic cocktail calculations.
While the p+p result is in good agreement with the cocktail calculation and a small 
excess over the cocktail calculation is observed for $p_{T} > 3.0~$GeV/$c$, an enhancement 
over the cocktail calculation is clearly seen in Au+Au collisions in 
$m_{ee} > 100~$MeV/$c^{2}$.

The following function, Eq.~\ref{eq_fit} is fitted to the data in order to determine the 
fraction of the direct $\gamma^{*}$ component in the $e^{+}e^{-}$ mass distribution.
\begin{equation}
f(m_{ee}) = (1-r) \cdot f_{cocktail}(m_{ee}) + r \cdot f_{direct}(m_{ee}),
\label{eq_fit}
\end{equation}
where $f_{cocktail}$ is the mass distribution from the decay of neutral hadrons estimated 
using the cocktail calculation and $f_{direct}$ is the expected distribution from the 
direct $\gamma^{*}$ decays~\cite{ppg086}, and $r$ is the direct $\gamma^{*}$ fraction.
The right panel in Fig.~\ref{mass_fit} shows the e$^{+}$e$^{-}$ mass distribution in 
Au+Au collisions for $1.0 < p_{T} < 1.5~$GeV/$c$ together with the fit result by 
Eq.~\ref{eq_fit} shown as a thick solid black line.
Assuming that this excess comes from direct $\gamma^{*}$ decays, the fit result agrees 
with the data even in $m_{ee} > 300~$MeV/$c^{2}$ shown as a dotted black line.
On the other hand, assuming that the excess comes from $\eta$, i.e. $f_{\eta}(m_{ee})$ is
used in place of $f_{direct}(m_{ee})$, anomalous $\eta$ enhancement is required to reproduce 
the excess.
Therefore the assumption that the excess comes from direct $\gamma^{*}$ decays seems 
to be favored by the data.
\begin{figure}[htbp]
  \centering
  \vspace{-0.35cm}
  \resizebox{14.25cm}{!}{\includegraphics{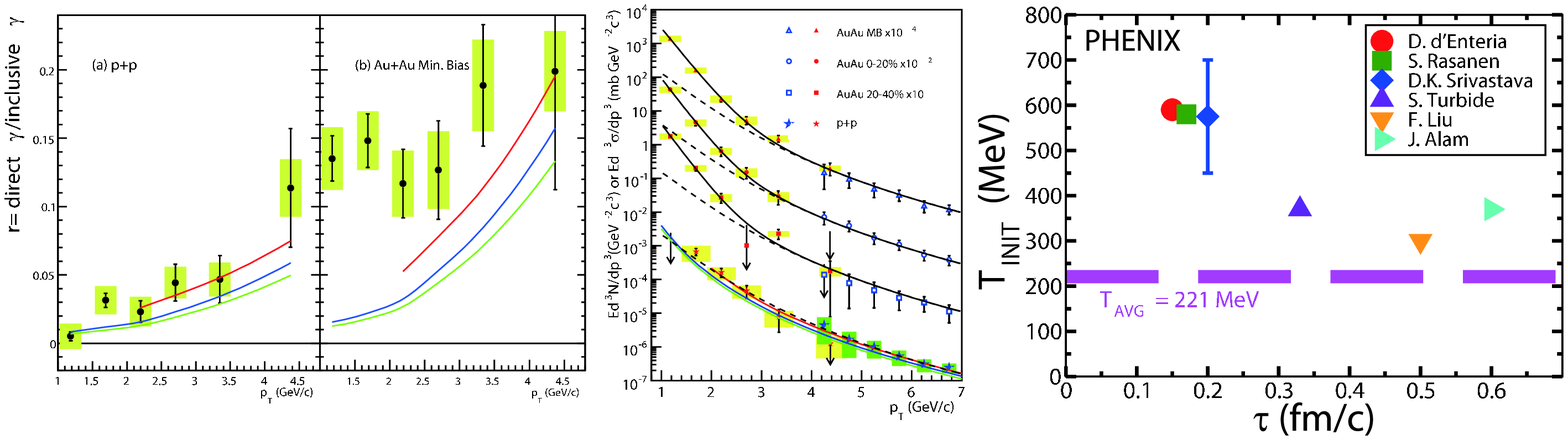}}
  \vspace{-0.5cm}
  \caption[]{The left panel shows the obtained direct $\gamma^{*}$ fractions as a function 
    of $p_{T}$ in p+p and Au+Au collisions. The center panel shows the direct photon spectra 
    in p+p and Au+Au collisions, and the right panel shows the relation between the initial 
    temperature and formation time in several hydrodynamical models which are in agreement 
    with the data.}
  \label{d_photon}
\end{figure}
The left panel in Fig.~\ref{d_photon} shows the obtained direct $\gamma^{*}$ fractions as 
a function of $p_{T}$ in p+p and Au+Au collisions.
The symbols show the results and the lines are the expectations from NLO pQCD calculations 
with different theoretical scales~\cite{NLOpQCD}.
A clear excess above the expectation is seen in Au+Au collisions while the p+p 
result is consistent with the expectation.
Finally, the real direct photon yields are converted from the direct $\gamma^{*}$ fractions.
The center panel in Fig.~\ref{d_photon} shows the direct photon spectra in p+p and Au+Au 
collisions.
The closed and open symbols show the results from 'virtual' and 'real' photon analyses, 
respectively.
The star symbol shows the p+p result and the triangle, circle and box symbols show the Au+Au 
results for minimum bias, centrality of 0-20\% and 20-40\%.
This is the first measurement where the direct photon yield in p+p collisions has been 
successfully measured in $1.0 < p_{T} < 3.0~$GeV/$c$.
The p+p result is consistent with NLO pQCD calculations in $p_{T} > 2.0~$GeV/$c$, but a 
modified power law function as shown by the dashed curve, $A_{pp}(1+p_{T}^{2}/B)^{-n}$, 
describes the p+p result better.
It serves as a crucial reference to the Au+Au result.
In Au+Au collisions, a significant excess over the binary-scaled p+p result is observed in 
$p_{T} <3.0~$GeV/$c$.
An exponential function, $A_{AuAu}exp(-p_{T}/T)$, plus the binary-scaled p+p result is fitted 
to the Au+Au result in order to determine the inverse slope, $T$ which is related to the 
temperature of the matter.
In central Au+Au collisions, the inverse slope of the fitting function is 
$221 \pm 23 \pm 18~$MeV.

Various efforts for understanding the implications of the result have been made via 
hydrodynamical approach~\cite{theo_comp}.
They are in agreement with the data within a factor of 2 in spite of their wide-ranging 
initial conditions, $300 < T_{INIT} < 600~$MeV and $0.15 < \tau < 0.5~$fm/$c$.
The right panel in Fig.~\ref{d_photon} shows the relation between the initial temperature and 
formation time in the hydrodynamical models.
These models indicate that the temperature of the matter at RHIC is higher than the critical 
temperature predicted by lattice QCD (160-190~MeV).

\section{Measurement in d+Au collisions}\label{vphoton_dAu}
The same measurement in d+Au collisions is also very important to evaluate the contribution 
of the nuclear effects such as Cronin effect and nuclear shadowing.
These effects may modify the photon yield in the low $p_{T}$ region.
The d+Au data already has been taken in 2003, and the direct photon measurement with the 
virtual photon method has been performed.
\begin{figure}[htbp]
  \centering
  \vspace{-0.3cm}
  \resizebox{9.30cm}{!}{\includegraphics{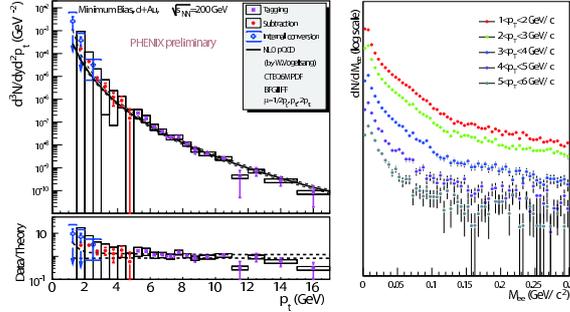}}
  \vspace{-0.5cm}
  \caption[]{The left panel shows the direct photon spectrum in d+Au collisions using the 
    data taken in 2003 together with the NLO pQCD calculation. The right panel shows the 
    $e^{+}e^{-}$ mass spectra in $m_{ee} < 300~$MeV/$c^{2}$ for several $p_{T}$ ranges 
    ($1.0 < p_{T} < 6.0~$GeV/$c$) using the data taken in 2008.}
  \label{photon_dAu}
\end{figure}
The left panel in Fig.~\ref{photon_dAu} shows the direct photon spectrum in d+Au collisions 
using the data taken in 2003~\cite{dAu}.
The result from the 'virtual' photon measurement could not reach to $p_{T} > 3.0~$GeV/$c$ 
due to low statistics.
The d+Au result seems to be consistent with the NLO pQCD calculation shown as a line, but the 
systematic uncertainties are still too large to allow a conclusion.
To evaluate the nuclear effect on the low $p_{T}$ photon yield surely, d+Au data have been 
taken again in 2008 with statistics about 30 times greater than that of the data taken in 2003.
The right panel in Fig.~\ref{photon_dAu} shows the $e^{+}e^{-}$ mass spectra in 
$m_{ee} < 300~$MeV for several $p_{T}$ ranges using the data taken in 2008.
The direct photon spectrum measured in these data is expected to reach 6.0~GeV/$c$.

\section{Summary and Outlook}\label{summary}
The direct photon measurements with the virtual photon method have been performed for p+p 
and Au+Au collisions at PHENIX.
This is the first measurement where the direct photon yield in p+p collisions has been
successfully measured in $1.0 < p_{T} < 3.0~$GeV/$c$.
A significant excess over the binary-scaled p+p result is observed in the direct photon 
yield in Au+Au collisions.
The excess is fitted with the exponential function and the inverse slope is given as 
$221 \pm 23 \pm 18~$MeV in central Au+Au collisions.
Several hydrodynamical models which reproduce the Au+Au result indicate that the temperature 
of the matter at RHIC is higher than the critical temperature of QGP.
The direct photon measurement in d+Au collisions has been made using the data taken in 2003 
to evaluate the contribution of the nuclear effects.
The d+Au result seems to be consistent with NLO pQCD calculation, but systematic 
uncertainties are too large due to small statistics.
The d+Au data taken in 2008 are expected to provide better result because its statistics 
is about 30 times more than that of the data taken in 2003.



\end{document}